\documentstyle[12pt]{article}
\begin{document}
\addtolength{\baselineskip}{0.5\baselineskip}

\rightline{CALT-68-1983}
\vskip 2cm
\begin{center}
{\large\bf Black Hole Solutions of Kaluza-Klein Supergravity Theories
and String Theory}
\end{center}
\vskip 1cm
\begin{center}
Jaemo Park\footnote{e-mail: jpk@theory.caltech.edu, jaemo@cco.caltech.edu}
\end{center}
\begin{center}
 {\em California Institute of Technology, Pasadena, CA 91125}
\end{center}
\vskip 3cm
\centerline{ABSTRACT}
\vskip 1cm
\begin{quote}
We find $U(1)_{E} \times U(1)_{M}$ non-extremal black hole solutions of
6-dimensional Kaluza-Klein
supergravity theories. Extremal solutions were found by  Cveti\v{c} and
Youm\cite{C-Y}. Multi black hole configurations are also presented. After
electro-magnetic duality
transformation is performed, these multi  black hole solutions are mapped
into the exact solutions found by Horowitz and Tseytlin\cite{H-T} in
5-dimensional string theory compactified into
4-dimensions. The massless fields of this theory can be embedded into
the heterotic
 string theory compactified on a 6-torus.
Rotating black hole solutions of this string theory
can be read off from those of the heterotic string theory found by
Sen\cite{Sen3}.
\end{quote}

\newpage
\section{\large\bf  Introduction}
There has been considerable interest in the connection between  black holes
and  supersymmetry. The familiar example is the
Reissner-Nordstr\"{o}m black
hole. This can be embedded in N=2 supergravity and the charge
of the black hole
is identified with the central charge of the extended supersymmetry
algebra\cite{Gibbons}. Furthermore, using methods similar to those
used to prove
 the positivity
of the ADM mass of a gravitational system\cite{WN}, it was shown that
the mass of the
black hole is bounded from below by its charge.\footnote{There are some
assumptions in proving this result. One of them is that the charge to mass
ratio
is less than or equal to 1 for any small volume of matter. Even with this
assumption it is nontrivial that  black hole mass is bounded below by
its charge, since  black holes can be formed through a complicated
gravitational collapse.} This is exactly the same bound
that must be satisfied for black holes to be free of naked singularities.
The extremal solutions saturate the mass bound and admit a Killing spinor
which is constant with respect to the supercovariant
derivative. This condition gives first order differential equations to be
satisfied by the extremal solutions, i.e., Killing
 spinor equations.

 Similar phenomena have been found for other cases as well.
One example is the charged black hole arising from  string
theory\cite{Garf}. One particular feature of  string theory is the
nonpolynomial coupling of a scalar
field to gauge fields. This is also a common characteristic of Kaluza-Klein
theories\cite{GM}. Those theories have terms like
$ e^{ 2\alpha\phi}F_{\mu\nu}F^{\mu\nu}$ in
the action where $F_{\mu\nu}$ is the field strength
of a gauge field. Charged
black hole solutions arising from such theories have
attracted much attention
because they have drastically different causal structures
and thermodynamic
properties than the Reissner-Nordstr\"{o}m black hole.
The $\alpha =1$ case is the string theory  and the supersymmetric properties
of the
 black hole solutions in this case were studied in detail
in ref.\cite{Kallosh}.
Also for $\alpha=\sqrt{3}$ the supersymmetric embedding is known and it is
5-dimensional Kaluza-Klein supergravity\cite{sq3}.
But it was conjectured that for an arbitrary value of $\alpha$ the
corresponding
black hole solution admits an embedding in some supergravity
theory\cite{Soliton}.

 One of the motivations of the paper by Cveti\v{c} and Youm\cite{C-Y},
is to find such embeddings for different values of $\alpha$ using the
dimensional
reduction of higher dimensional supergravity theories. They started with the
minimal supersymmetric extension of pure gravity in $(4+n)$ dimensions.
Keeping only the pure gravity part and performing the dimensional
reduction, they obtained the 4-dimensional theory with two Abelian gauge
fields.   It turns out that the resulting bosonic action for each $n$ can be
reduced to the action obtained from the 6-d Kaluza-Klein supergravity.
 They found the extremal black hole
configurations using the Killing spinor equations.

 One purpose of this paper is to find the black hole solutions of the 6-d
Kaluza-Klein supergravity by directly
solving the equations of motion, thereby obtaining non-extremal black hole
solutions as well. This is presented in section 2. Global structures and
thermal properties of the black hole solutions are explained.
It turns out that the black hole solutions are intimately related with black
hole solutions in string theory. We devote section 3 and section 4 to
discussing those connections. In section 3 we present multi black hole
solutions of the 6-d Kaluza-Klein supergravity theories. After
electro-magnetic duality transformation is made, these solutions are mapped
into the exact solutions of 5-d string theory compactified into 4-dimensions.
This theory is considered by Horowitz and Tseytlin\cite{H-T}. We briefly
discuss 5-d geometry of the exact solutions.
In section 4 we show that after a field redefinition, the massless fields
of 5-d string theory can be embedded into the heterotic string theory.
The general electrically charged, rotating black hole solutions are studied
by Sen\cite{Sen3}. Thus we can read off the black hole solutions of the 5-d
string
theory from those of the heterotic string theory. In this way, we obtain
the rotating black hole solutions of the 5-d string theory compactified into
4-dimensions.
Conclusions and speculations are presented in section 5.

\section{\large\bf  Black hole solutions of the 6-D Kaluza-Klein theory}
 The bosonic action in 4+n-dimension is of the  form
\footnote{With regard to the metric sign and the definition of the curvature,
we follow the convention used by Misner, Thorne and Wheeler\cite{MTW}. The
metric signature is $(-+\cdots +)$, $R^{\alpha}_{\, \beta\gamma\delta} =
\partial_{\gamma}\Gamma^{\alpha}_{\,\beta\gamma\delta} + \cdots$,
$R_{\mu\nu}=R^{\alpha}_{\,\mu\alpha\nu}$.}
\begin{equation}
S_{4+n} = \frac{1}{16\pi G_{4+n}} \int \sqrt{-g^{(4+n})}\,  d^{4+n}x\,
  R^{(4+n)}.
\end{equation}
$G_{4+n}$ is the gravitational constant in $4+n$-dimension.
The higher dimensional metric $ g_{AB}^{(4+n)}$ is taken as
\begin{equation}
g_{AB}^{(4+n)} = \left( \begin{array}{cc}
 e^{-\frac{1}{\alpha}\psi}g_{\lambda\pi}
+ e^{\frac{2\psi}{n\alpha}} \rho_{\tilde{\lambda}\tilde{\pi}}
A_{\lambda}^{\tilde{\lambda}} A_{\pi}^{\tilde{\pi}}
&   e^{\frac{2\psi}{n\alpha}}\rho_{\tilde{\lambda}\tilde{\pi}}
A_{\lambda}^{\tilde{\lambda}}  \\  &  \\
  e^{\frac{2\psi}{n\alpha}} \rho_{\tilde{\lambda}\tilde{\pi}}
A_{\pi}^{\tilde{\pi}}   &
  e^{\frac{2\psi}{n\alpha}} \rho_{\tilde{\lambda}\tilde{\pi}}
 \end{array}  \right).
\end{equation}
Greek lower-case letters denote the space-time indices in 4-dimension while
lower-case letters with tilde are used for the internal coordinates.
 $ \rho_{\tilde{\lambda}\tilde{\pi}}$ satisfies
det$\rho_{\tilde{\lambda}\tilde{\pi}} =1$, i.e.,
$ \rho_{\tilde{\lambda}\tilde{\pi}}$ is the unimodular part of the
internal metric. All fields have no dependence on the internal
coordinates.
It is shown in ref. \cite{C-Y} that if all the gauge fields are Abelian, the
supersymmetric configuration is possible only if the the internal group
$U(1)^{n}$ is broken down to $U(1)_{E} \times U(1)_{M}$.
Thus electric charge and
magnetic charge should be associated to different $U(1)$ sectors.
The internal metric $\rho_{\tilde{\alpha}\tilde{\beta}}$ is assumed to be
\begin{equation}
\rho_{\tilde{\alpha}\tilde{\beta}} = {\rm diag}(\rho_{1}, \cdots, \rho_{n-1},
\prod_{k=1}^{n-1}\rho_{k}) .
\end{equation}

 Then the resulting action in 4-dimensions after the trivial integration over
the internal coordinates is
\begin{eqnarray}
S_{4} &=& \frac{1}{16\pi G} \int \sqrt{-g}\,  d^4 x\,  ( R
-\frac{1}{4}e^{\alpha\psi} \rho_{n-1} (F_{\mu\nu}^{n-1})^2
-\frac{1}{4}e^{\alpha\psi} \rho_{n} (F_{\mu\nu}^{n})^2  \\
 & & -\frac{1}{2}\partial_{\mu}\psi\partial^{\mu}\psi
-\frac{1}{4}\sum_{i=1}^{n}\partial_{\mu}\log
\rho_{i}\partial^{\mu} \log \rho_{i} ),  \nonumber
\end{eqnarray}
where $\alpha = \sqrt{\frac{n+2}{n}}$. The gravitational coupling constant $G$
in 4-dimension is given by $ G = G_{n+4}\, (2\pi R)^{n}$ for a
toroidal compactification where each internal dimension has radius R,
$F_{\mu\nu}^{n-1}$ and $F_{\mu\nu}^{n}$ denote the unbroken Abelian
gauge group.
This action can be reduced to that
of 6-d Kaluza-Klein theory with the  field redefinition,
\begin{eqnarray}
\phi \equiv\frac{1}{\sqrt{2}\alpha}\,\psi \quad \quad ,
\chi_{i} \equiv \frac{1}{2\sqrt{2}} (\log \rho_{i} + \frac{2}{n\alpha}\psi)
\quad ,     i = 1,\cdots, n-2 \nonumber \\
\chi_{n-1} \equiv \frac{1}{2\sqrt{2}} (\log \rho_{n-1}
+ \frac{2-n}{n\alpha}\psi), \quad
\chi_{n} \equiv \frac{1}{2\sqrt{2}} (\log \rho_{n} + \frac{2-n}{n\alpha}\psi).
 \label{eq:field}
\end{eqnarray}
Then the corresponding action is
\begin{eqnarray}
S_{4} &=& \frac{1}{16\pi G} \int \sqrt{-g}\, d^4x\, ( R
- \frac{1}{4} e^{2\sqrt{2}(\phi+\chi_{n-1})} ( F_{n-1} )^2
- \frac{1}{4} e^{2\sqrt{2}(\phi+\chi_{n})} ( F_{n} )^2  \\
   & &   -2 \sum_{i=1}^{n} (\nabla \chi_{i})^2
 -2 \partial_{\mu} \phi ( \partial^{\mu} \chi_{n-1}
+ \partial^{\mu} \chi_{n} )
 -2 (\nabla \phi)^2 ). \nonumber
\end{eqnarray}
We see that the field equations derived from this action will admit a
solution with $\chi_{i}$  set to constant for $i = 1,\cdots, n-2$. This
implies that $ \chi_{n-1} + \chi_{n}$ is also constant since
$\sum_{i=1}^{n} \chi_{i} = 0$. Absorbing such constants into field
redefinition of the gauge fields and defining
$\chi\equiv \chi_{n-1}, K_{\mu\nu} \equiv F^{n-1}_{\mu\nu},
F_{\mu\nu} \equiv F^{n}_{\mu\nu}$ , we finally obtain the following action
\footnote{ From now on we set $G = c = 1$.}
\begin{equation}
S=\frac{1}{16\pi} \int \sqrt{-g}\, d^4x\, (R- 2 (\nabla \phi)^2
- 4 (\nabla \chi)^2
 -e^{2\sqrt{2}(\phi+\chi)} K^2 -e^{2\sqrt{2}(\phi-\chi)} F^2) .
\label{eq:action}
\end{equation}

If the metric in 4-dimensions takes the form
\begin{equation}
  ds^2 = -A^2(r) dt^2 + \frac{dr^2}{A^2(r)} + R^2(r)  d\Omega^2 ,
\end{equation}
the equations for the gauge fields are solved by
\begin{equation}
K_{\theta\phi}= Q_{m} \sin\theta, \quad
F^{rt} = \frac{Q_{e}}{R^2 e^{2\sqrt{2}(\phi-\chi)}}.
\end{equation}
Then the  equations of the two scalar fields are
\begin{equation}
\frac{1}{R^2}\frac{d}{dr} (R^2 A^2 \frac{d\phi}{dr}) =
\sqrt{2} e^{2\sqrt{2}(\phi+\chi)} \frac{Q_{m}^2}{R^4}
 -\sqrt{2} e^{-2\sqrt{2}(\phi-\chi)} \frac{Q_{e}^2}{R^4},  \label{eq:S1}
\end{equation}
\begin{equation}
\frac{1}{R^2} \frac{d}{dr} (R^2 A^2 \frac{d\chi}{dr})=
\frac{\sqrt{2}}{2} e^{2\sqrt{2} (\phi+\chi)} \frac{Q_{m}^2}{R^4}
 +\frac{\sqrt{2}}{2} e^{-2\sqrt{2} (\phi-\chi)} \frac{Q_{e}^2}{R^4}.
 \label{eq:S2}
\end{equation}
The gravitational field equations are
\begin{eqnarray}
 G^{t}_{t}&=& \frac{A^2 (R^{'})^2 -1}{R^2} + \,
\frac{2 A^2 R^{''} + 2 A A^{'} R^{'}}{R} \label{eq:gtt} \\
         &=&-A^2 (\frac{d\phi}{dr})^2-2 A^2 (\frac{d\chi}{dr})^2
 -e^{2\sqrt{2}(\phi+\chi)}\frac{Q_{m}^2}{R^4}
 -e^{-2\sqrt{2}(\phi-\chi)}\frac{Q_{e}^2}{R^4},  \nonumber
\end{eqnarray}
\begin{eqnarray}
 G^{r}_{r}&=& \frac{A^2 (R^{'})^2 -1}{R^2} +\,\frac{2 AA{'}R{'}}{R}
\label{eq:grr} \\
         &=& A^2 (\frac{d\phi}{dr})^2 + 2 A^2 (\frac{d\chi}{dr})^2
 -e^{2\sqrt{2}(\phi+\chi)}\frac{Q_{m}^2}{R^4}
 -e^{-2\sqrt{2}(\phi-\chi)}\frac{Q_{e}^2}{R^4}, \nonumber
\end{eqnarray}
\begin{eqnarray}
G_{\theta}^{\theta}
  &=&\frac{1}{2} (A^2)^{''} + \frac{2 A A^{'} R^{'}}{R}
+\frac{A^2 R{''}}{R}  \\
        &=&-A^2 (\frac{d\phi}{dr})^2 - 2 A^2 (\frac{d\chi}{dr})^2
 +e^{2\sqrt{2}(\phi+\chi)}\frac{Q_{m}^2}{R^4}
 +e^{-2\sqrt{2}(\phi-\chi)}\frac{Q_{e}^2}{R^4}, \nonumber
\end{eqnarray}
where $'$ means the differentiation by r.
 From these equations, we obtain
\begin{equation}
R^2 (G_{r}^{r} + G_{\theta}^{\theta}) = \frac{1}{2} (A^2 R^2)^{''} -1 = 0 ,
\end{equation}
\begin{eqnarray}
R^2 (2 G_{\theta}^{\theta} + G_{r}^{r} - G_{t}^{t}) & &  \nonumber  \\
 =((A^2)^{'} R^2)^{'}&=& 2 e^{2\sqrt{2}(\phi+\chi)}\frac{Q_{m}^2}{R^4}
 +2 e^{-2\sqrt{2}(\phi-\chi)}\frac{Q_{e}^2}{R^4} .  \label{eq:A1}
\end{eqnarray}
Since we are searching for solutions with a regular horizon,  we can set
 $A^2 R^2 =(r-r_{+}) (r-r_{-})$ with $ r_{+} \geq r_{-} $ where
$r=r_{+}$ is  the location of the horizon.
If we define
\begin{equation}
\rho \equiv \frac{1}{r_{+}-r_{-}}
\log (\frac{r-r_{+}}{r-r_{-}}),
\end{equation}
 then we can write
\begin{equation}
 A^2 R^2 \frac{d}{dr} = \frac{d}{d\rho}. \nonumber
\end{equation}
Throughout this section $\rho$ is used as a short notation for the more
complicated expression of r.
Comparing the eq.~(\ref{eq:S2}) and eq.~(\ref{eq:A1}) we have
\begin{equation}
\frac{d^2}{d\rho^2}2\sqrt{2}\chi = \frac{d^2}{d\rho^2} \log  A^2 , \nonumber
\end{equation}
\begin{equation}
2\sqrt{2}\chi = \log  A^2 + \beta \rho + 2\sqrt{2} \chi_{\infty} .
\end{equation}
Here $\beta$ is a constant and $\chi_{\infty}$ is the asymptotic value of
$\chi$. We require that $\chi$ be regular and finite at the horizon. Thus
$\beta = -(r_{+}-r_{-})$ and
\begin{equation}
e^{2\sqrt{2}\chi} = A^2\, \frac{r-r{-}}{r-r{+}}\, e^{2\sqrt{2}\chi_{\infty}},
 \label{eq:A2}
\end{equation}
\begin{equation}
R^2 e^{2\sqrt{2}\chi} = (r-r_{-})^2 e^{2\sqrt{2}\chi_{\infty}}, \label{eq:R2}
\end{equation}
\begin{equation}
\frac{R^{''}}{R} =
-\frac{2\sqrt{2}\chi^{'}}{r-r_{-}} -\sqrt{2} \chi^{''} +
2(\chi^{'})^2 \label{eq:R}.
\end{equation}
Subtracting ~(\ref{eq:grr}) from ~(\ref{eq:gtt}), we obtain
\begin{equation}
 \frac{R^{''}}{R} =
- (\frac{d\phi}{dr})^2 - 2 (\frac{d\chi}{dr})^2 . \label{eq:1}
\end{equation}
Now the equations of the scalar fields are written as
\begin{equation}
\frac{d^2}{d\rho^2}(2\sqrt{2}\phi) = 4 Q_{m}^2 A^2 e^{2\sqrt{2}(\phi+\chi)}
 - 4 Q_{e}^2 A^2 e^{-2\sqrt{2}(\phi-\chi)},   \label{eq:2}
\end{equation}
\begin{equation}
\frac{d^2}{d\rho^2}(4\sqrt{2}\chi) = 4 Q_{m}^2 A^2 e^{2\sqrt{2}(\phi+\chi)}
 + 4 Q_{e}^2 A^2 e^{-2\sqrt{2}(\phi-\chi)}.   \label{eq:3}
\end{equation}
Eq.~(\ref{eq:1}), ~(\ref{eq:2}), ~(\ref{eq:3}) are the main equations
to be considered.

 If $Q_{e} = 0$, we see that $\phi-\phi_{\infty} = 2 \chi - 2\chi_{\infty}$.
Using this relation and eq.~(\ref{eq:R}),  eq.~(\ref{eq:1}) is easily
integrated to give
\begin{eqnarray}
e^{2\sqrt{2}\chi}&=& e^{2\sqrt{2}\chi_{\infty}}\sqrt{\frac{r-r_{-}}{r+B}} ,
\label{eq:sol1} \\
e^{2\sqrt{2}\phi}&=& e^{2\sqrt{2}\phi_{\infty}}\frac{r-r_{-}}{r+B}.  \nonumber
\end{eqnarray}
{}From eq.~(\ref{eq:A2}) and eq.~(\ref{eq:R2}) we find
\begin{eqnarray}
A^2 &=& \frac{r-r_{+}}{\sqrt{(r+B)(r-r_{-})}} ,\label{eq:sol2} \\
R^2 &=& (r-r_{-}) \sqrt{(r+B)(r-r_{-})} . \nonumber
\end{eqnarray}
{}From eq.~(\ref{eq:2}) we can see that $ (B+r_{+})(B+r_{-})=
4Q_{m}^2 e^{2\sqrt{2}(\phi_{\infty}+\chi_{\infty})}$.
We still have freedom in r-coordinate choice $r \rightarrow r+\Gamma$,
using this we can set B=0. Since $\phi$ is proportional to $\chi$ up to
constant, we can define a new field $\phi^{\prime}$ as
$ -2\sqrt{3}\phi^{'} \equiv 2\sqrt{2}(\phi +\chi)$. The action
in terms of $\phi^{\prime}$ instead of $\phi$ and $\chi$ for $Q_{e}=0$ case is
\begin{equation}
 S = \frac{1}{16\pi} \int \sqrt{-g} d^4 x
( R - 2 (\nabla \phi^{'})^2 - e^{-2\sqrt{3} \phi^{'}} K^2 ). \label{eq:rt3}
\end{equation}
This is the action with  $\alpha=\sqrt{3}$ for the Maxwell-dilaton coupling
 considered in \cite{Garf}.
Furthermore one can see that the known solution of ~(\ref{eq:rt3}) coincides
with the above. Similar analysis can be performed for a purely electric case.
 As stated before, $\alpha=\sqrt{3}$ case corresponds to the 5-dimensional
 Kaluza-Klein theory, and we conclude that  the solutions of a purely
electric or purely magnetic case
are those of 5-dimensional
Kaluza-Klein supergravity theories.
Also from the above analysis we know that the solution of
\begin{eqnarray}
\frac{d^2}{d\rho^2} \psi = P^2\, \frac{r-r_{+}}{r-r_{-}}\, e^{2\psi}
\label{eq:n1}
\end{eqnarray}
for constant P is given by
\begin{eqnarray}
e^{\psi} = \frac{r-r_{-}}{r+B}, \quad  (B + r_{+})(B +r_{-}) = P^2.
\label{eq:n2}
\end{eqnarray}

 If we consider the general case where both electric charge and magnetic
charge are nonzero, eq.~(\ref{eq:2}) and eq.~(\ref{eq:3}) can be written as
\begin{eqnarray}
\frac{d^2}{d\rho^2} (\frac{2\sqrt{2}\phi + 4\sqrt{2}\chi}{2})
&=& 4 Q_{m}^2 e^{-2\sqrt{2}\chi_{\infty}} \frac{r-r_{+}}{r-r_{-}}
 e^{2\sqrt{2}\phi + 4\sqrt{2}\chi},  \\
\frac{d^2}{d\rho^2} (\frac{-2\sqrt{2}\phi + 4\sqrt{2}\chi}{2})
&=& 4 Q_{e}^2 e^{-2\sqrt{2}\chi_{\infty}} \frac{r-r_{+}}{r-r_{-}}
 e^{-2\sqrt{2}\phi + 4\sqrt{2}\chi}.
\end{eqnarray}
These equations have the same form as ~(\ref{eq:n1}) and this suggests that
solutions have the form of ~(\ref{eq:n2}). Hence
\begin{eqnarray}
\exp \frac{2\sqrt{2}(\phi-\phi_{\infty}) + 4\sqrt{2}(\chi -\chi_{\infty})}{2}
 &=& \frac{r-r_{-}}{r+B}, \nonumber\\  & & \nonumber \\
\exp \frac{-2\sqrt{2}(\phi-\phi_{\infty}) + 4\sqrt{2}(\chi -\chi_{\infty})}{2}
 &=& \frac{r-r_{-}}{r+C}. \nonumber
\end{eqnarray}
We can choose $C = -B = r_{0}$ by a suitable shift
of $r \rightarrow r + \Gamma$.
It is easily checked that this ansatz works.
Thus the black hole solutions are given by
\begin{eqnarray}
A^2 &=& \frac{r-r_{+}}{\sqrt{r^2 - r_{0}^2}}, \label{eq:bl}\\
R^2 &=& (r-r_{-}) \sqrt{r^2 - r_{0}^2}, \nonumber \\
e^{2\sqrt{2}\phi} &=&
e^{2\sqrt{2}\phi_{\infty}} \frac{r+r_{0}}{r-r_{0}}, \nonumber \\
e^{2\sqrt{2}\chi}
&=& e^{2\sqrt{2}\chi_{\infty}} \frac{r-r_{-}}{\sqrt{r^2-r_{0}^2}},
\nonumber \\
F^{rt}&=& \frac{Q_{e}}{(r+r_{0})^{2}}
e^{-2\sqrt{2}(\phi_{\infty}-\chi_{\infty})} , \nonumber \\
K_{\theta\phi}&=& Q_{m} \sin\theta .  \nonumber
\end{eqnarray}
And the constraints on $ r_{+}, r_{-}, r_{0} $ are
\begin{eqnarray}
 ( r_{+}+r_{0})( r_{-}+r_{0} ) &=& 4 Q^2,  \label{eq:r} \\
 ( r_{+}-r_{0})( r_{-}-r_{0} ) &=& 4 P^2 .  \nonumber
\end{eqnarray}
Here $P$ and $Q$ are defined as
\begin{eqnarray}
Q^2 &\equiv& Q_{e}^2 e^{-2\sqrt{2} \phi_{\infty}
+ 2\sqrt{2} \chi_{\infty}}, \\
P^2 &\equiv& Q_{m}^2 e^{2\sqrt{2} \phi_{\infty} + 2\sqrt{2} \chi_{\infty}}.
\nonumber
\end{eqnarray}

 From solution ~(\ref{eq:bl}),  we see that $ r_{+} = 2 M $
where $M$ is the mass of
a black hole,  $  r_{0} = \sqrt{2}\Sigma $ where $ \Sigma $ is the charge of
the scalar $\phi$, and $r_{-} = -2 \Delta$ where $\Delta$ is the charge of
$\chi$ . The charge $\Sigma$ is defined by
\begin{equation}
\phi = \phi_{\infty} + \frac{\Sigma}{r} +O(\frac{1}{r^2}),\quad
r \rightarrow \infty.
\end{equation}
On the other hand, $\Delta$  is defined by
\begin{equation}
\sqrt{2}\chi = \sqrt{2}\chi_{\infty} + \frac{\Delta}{r} +O(\frac{1}{r^2}),
\quad  r \rightarrow \infty ,
\end{equation}
 since we adopt the different normalization for $\phi$ and $\chi$ in
action ~(\ref{eq:action}). Clearly $\Sigma$ and $\Delta$ are not independent
parameters and depend on $M, P$ and $Q$. Their dependence is given by cubic
equations but those equations are not particularly illuminating.
  Since $P$ and $Q$ depend on $Q_{e}$, $Q_{m}$ and $\phi_{\infty}$,
$\chi_{\infty}$ in turn, the black hole solutions are characterized
by five parameters i.e., mass, electric charge, magnetic charge, and
asymptotic values of scalar fields. From eq.~(\ref{eq:r}), one can easily
find that
$ M  \geq \frac{|Q|+|P|}{2}$ and
$ |\Sigma| \leq \frac{||Q|-|P||}{\sqrt{2}}$. The sign of $\Sigma$ is the same
as that of $|Q| - |P|$. Also $r_{+}\geq r_{-} \geq |r_{0}|$.
Calculation of the curvature tensors indicates that $r=r_{+}$ is indeed a
regular horizon and $r = r_{-}$ is a curvature singularity. Comparing with
the Reissner-Nordstr\"{o}m black hole, the would-be inner horizon turns
into the
singularity. The extremal solutions where $r_{+}$ coincides with $r_{-}$
agree with
 the solutions found by Cveti\v{c} and Youm\cite{C-Y}. Note that
$ M = \frac{|Q| + |P|}{2}$, $\Sigma = \frac{|Q| - |P|}{\sqrt{2}}$ and
$ \Delta = - M $ so that $ M^2 + \Sigma^2 + \Delta^2 = Q^2 + P^2 $ in
extremal case. This is the force balance condition, as we will see later.

The causal structure of the non-extremal case is that of the
Schwarzschild black hole.
For extremal
 black holes, the situation is two-fold. If both electric and magnetic
charge are non-zero, the corresponding black hole has a null singularity.
This can be seen from the fact that the
radial null geodesics satisfy $ \pm dt\propto dr/(r-r_{+})$, which implies
that as $r \rightarrow r_{+}$, the geodesics reach arbitrarily large values of
$|t|$. This shows that an outgoing null geodesic must cross every ingoing null
geodesic. However, if either of the charges is zero, the singularity becomes
timelike naked. In this case $r_{+}=r_{-}=\pm r_{0}$ and $\pm dt \propto
dr/\sqrt{r-r_{+}}$ for radial null rays near $r=r_{+}$. The Penrose diagrams
are given for each case in Figure 1.

 The temperature of the black holes can be found from the periodicity
of its Euclidean continuation\cite{GiHw}, or alternatively from its surface
gravity. It is $T = \frac{1}{4\pi\sqrt{r_{+}^2 - r_{0}^2}}$. In the
extremal case,
$ r_{+} = |Q| + |P|$ and $r_{-} = |Q| - |P|$.
Hence $ T$ approaches $ \frac{1}{8\pi\sqrt{|P||Q|}}$ in the extremal limit.
The entropy of the black holes can
be evaluated in two ways. One may integrate the first law of thermodynamics.
Or one can calculate the thermodynamic functions directly, using the saddle
point approximation for the action of the black holes in the Euclidean
continuation\cite{GiHw}. Either way gives
$ S = \frac{1}{4}A = \pi (r_{+} -r_{-})\sqrt{r_{+}^2 -r_{0}^2}  $.
In the extremal limit, the black holes approach zero entropy and nonzero
temperature configurations.\footnote{There are claims that the extremal
black holes always have zero entropy and can be in equilibrium with thermal
 radiation at any temperature\cite{Hawking4}. This indicates that we have to
distinguish
between an extremal black hole and limiting configuration from a non-extremal
black hole, if the claim is right. }

So far we have presented the black hole solutions of 6-d Kaluza-Klein
supergravity. For other dimensions we can read off the result from
eq.~(\ref{eq:field}). However the same metric remains a solution for all
dimensions.

\section{\large\bf  Multi black hole solutions
and their string interpretation}

  Now we will look for a static solution representing a collection of
extremal black holes,
 with the following ansatz for the metric in isotropic coordinates
\begin{equation}
ds^2= -e^{2U} dt^2 + e^{-2U} ( dx_{i})^2 .
\end{equation}
The nonzero components of the Ricci tensor in the coordinate basis are
\begin{eqnarray}
R_{00} &=& e^{4U} \partial_{i}\partial_{i} U \\
R_{ij} &=& -2 \partial_{i} U \partial_{j} U
+ \delta_{ij} \partial_{k}\partial_{k} U  . \nonumber
\end{eqnarray}
If we choose the gauge fields as
\begin{equation}
F_{i0} = \partial_{i} \Psi,  \quad
K_{jk} = \varepsilon_{ijk} e^{-2\sqrt{2}(\phi+\chi) - 2U} \partial_{k}\lambda ,
\end{equation}
the equations of motion and Bianchi identities give
\begin{eqnarray}
\partial_{i} ( e^{2\sqrt{2}(\phi-\chi)-2U} \partial_{i}\Psi) &=& 0 , \\
\partial_{i} ( e^{-2\sqrt{2}(\phi+\chi)-2U} \partial_{i}\lambda) &=& 0 .
\nonumber
\end{eqnarray}
And the scalar field equations are
\begin{eqnarray}
\partial_{i}\partial_{i} \phi &=&
\sqrt{2} \,e^{-2\sqrt{2}(\phi+\chi)-2U} (\partial_{i} \lambda)^2
-\sqrt{2}\, e^{2\sqrt{2}(\phi-\chi)-2U} (\partial_{i} \Psi)^2 , \\
\partial_{i}\partial_{i} \chi &=&
\frac{\sqrt{2}}{2}\, e^{-2\sqrt{2}(\phi+\chi)-2U} (\partial_{i} \lambda)^2
+\frac{\sqrt{2}}{2}\, e^{2\sqrt{2}(\phi-\chi)-2U} (\partial_{i} \Psi)^2.
\nonumber
\end{eqnarray}
The gravitational field equations give
\begin{eqnarray}
\partial_{i}\partial_{i} U &=&
 e^{-2\sqrt{2}(\phi+\chi)-2U} (\partial_{i} \lambda)^2
+e^{2\sqrt{2}(\phi-\chi)-2U} (\partial_{i} \Psi)^2 , \\
\partial_{i}U\partial_{j}U &=& -\partial_{i}\phi \partial_{j}\phi -
2 \partial_{i}\chi \partial_{j}\chi +
 e^{-2\sqrt{2}(\phi+\chi)-2U} \partial_{i} \lambda \partial_{j} \lambda
+  e^{2\sqrt{2}(\phi-\chi)-2U} \partial_{i} \Psi \partial_{j} \Psi.
\end{eqnarray}
The relevant solutions of these equations are
\begin{eqnarray}
\sqrt{2}\chi = U +\sqrt{2}\chi_{\infty},  \label{eq:m1}  \\
e^{-\sqrt{2}(\phi+2\chi)} = H_{1}, \quad  e^{-\sqrt{2}(-\phi+2\chi)} = H_{2}
\nonumber, \\
2 e^{\sqrt{2}\chi_{\infty}} \lambda = \pm \frac{1}{H_{1}}, \quad
2 e^{\sqrt{2}\chi_{\infty}} \Psi = \pm \frac{1}{H_{2}},  \nonumber \\
\partial_{i}\partial_{i} H_{1} = 0,  \quad
\partial_{i}\partial_{i} H_{2} = 0 .
\nonumber
\end{eqnarray}

One particular solution representing the multiblack hole configuration is
given by
\begin{eqnarray}
H_{1}& =& e^{-\sqrt{2}(\phi_{\infty}+2\chi_{\infty})}
( 1 + \sum_{i=1}^{n} \frac{2|P_{i}|}{|x-x_{i}|}),  \label{eq:m2} \\
H_{2}& =& e^{\sqrt{2}(\phi_{\infty}-2\chi_{\infty})}
( 1 + \sum_{i=1}^{n} \frac{2|Q_{i}|}{|x-x_{i}|}). \nonumber
\end{eqnarray}
One can easily see that this is the collection of $n$ extremal black holes.
The relation between the parameters of each black hole is
\begin{eqnarray}
 M_{i} = -\Delta_{i} = \frac{|Q_{i}|+|P|_{i}}{2} , \label{eq:force} \\
\Sigma_{i} =\frac{|Q_{i}|-|P_{i}|}{\sqrt{2}}. \nonumber
\end{eqnarray}
This condition implies the force balance.
 To see this explicitly, let us consider gravitational,
electromagnetic, and scalar forces. The force between two distant
objects of masses and charges $( M_{1}, Q_{1}, P_{1}, \Sigma_{1}, \Delta_{1} )$
 and $( M_{2}, Q_{2}, P_{2}, \Sigma_{2}, \Delta_{2} )$ is given by
\begin{equation}
F_{12} = -\frac{M_{1} M_{2}}{ r_{12}^2} + \frac{ Q_{1} Q_{2}}{ r_{12}^2}
+ \frac{ P_{1} P_{2}}{ r_{12}^2} - \frac{ \Sigma_{1} \Sigma_{2}}{ r_{12}^2}
-\frac{ \Delta_{1} \Delta_{2}}{ r_{12}^2} .
\end{equation}
The scalar forces are attractive for charges of the same type and repulsive
for charges of opposite sign.
Using eq.~(\ref{eq:force}), we see that $F_{12}$ vanishes. This force
balance allows the black holes to be located at any place, in equilibrium
 with the other black holes.

Interestingly enough, the above solutions are related to the exact string
solutions found by Horowitz and Tseytlin\cite{H-T} by the electromagnetic
dual transformation.
{}From the action ~(\ref{eq:action}), if we perform the duality transformation
\begin{equation}
K_{\mu\nu}=\frac{1}{2} \varepsilon_{\mu\nu\lambda\sigma}C^{\lambda\sigma}
e^{-2\sqrt{2}(\phi+\chi)},
\end{equation}
the resulting action is
\begin{equation}
S_{1}=\frac{1}{16\pi} \int \sqrt{-g}\, d^4x\, (R- 2 (\nabla \phi)^2
- 4 (\nabla \chi)^2
 -e^{-2\sqrt{2}(\phi+\chi)} C^2 -e^{2\sqrt{2}(\phi-\chi)} F^2) ,
\end{equation}
 where $ \varepsilon_{\mu\nu\lambda\sigma}$ is an antisymmetric tensor with
$\varepsilon_{1234}=1$.
Now if we define
\begin{eqnarray}
F_{\mu\nu}^{s}  \equiv 2F_{\mu\nu}, & B_{\mu\nu}^{s} \equiv 2C_{\mu\nu} ,
\label{rel:1} \\
\varphi \equiv 2\sqrt{2} \chi ,  & \sigma \equiv \sqrt{2} \phi , \nonumber
\end{eqnarray}
the action is written as
\begin{equation}
S_{2}=\frac{1}{16\pi} \int \sqrt{-g}\, d^4x\, (R- (\nabla \sigma)^2
 - \frac{1}{2} (\nabla \varphi)^2
 -\frac{1}{4}e^{-2\sigma-\varphi} (B_{\mu\nu}^{s})^2
-\frac{1}{4}e^{2\sigma-\varphi} (F_{\mu\nu}^{s})^2).
\end{equation}
This is the dimensional reduction of the 5-d bosonic string action.
To see this, we start with the leading-order term in 5-d bosonic string action
\begin{equation}
S_{5}=\kappa^{0}\int d^5x\sqrt{-g_{(5)}}e^{-2\phi_{s}}(R+4(\nabla \phi_{s})^2
-\frac{1}{12}(H_{MNK})^2+o(\alpha^{\prime})) , \label{eq:SS6}
\end{equation}
where $H_{MNK}=3\partial_{[M}B^{\prime}_{NK]}=\partial_{M}B^{\prime}_{NK}
+\partial_{K}B^{\prime}_{MN}+\partial_{N}B^{\prime}_{KM}$. Here
$\partial_{[M}B^{\prime}_{NK]}$ is an  antisymmetric third rank tensor
of strength 1.\footnote{From now on we denote an antisymmetric $n$-th rank
tensor
of strength 1
as $ A_{[\mu_{1}\mu_{2} \cdots \mu_{n}]}$.}
Assuming that all the fields are independent of $x^5$, we obtain the 4-d
reduced action
\begin{eqnarray}
\tilde{S_{4}}&=&\hat{\kappa}_{0} \int d^4x\sqrt{{-g}}
e^{-2\phi_{s}+\sigma}(\hat{R}
+4(\partial_{\mu}\phi_{s})^2-4\partial_{\mu}\phi_{s}\partial^{\mu}\sigma
\label{eq:n3}\\
 & &-\frac{1}{12}(\hat{H}_{\mu\nu\lambda})^2-\frac{1}{4}e^{2\sigma}
(F_{\mu\nu}^{s})^2-\frac{1}{4}e^{-2\sigma}(B_{\mu\nu}^{s})^2+
o(\alpha^{\prime})) \nonumber
\end{eqnarray}
where
\begin{eqnarray}
 g_{55} \equiv e^{2\sigma}, & F_{\mu\nu}^{s}=\partial_{\mu}A^{s}_{\nu}-
\partial_{\nu}A^{s}_{\mu},  \\
 B_{\mu\nu}^{s}=\partial_{\mu}B^{s}_{\nu}-\partial_{\nu}B^{s}_{\mu},
&A_{\mu}^{s} \equiv g_{\mu 5}e^{-2\sigma} , \nonumber \\
B^{s}_{\mu} \equiv B^{\prime}_{\mu 5}, & \hat{H}_{\lambda\mu\nu}=
3\partial_{[\lambda}B^{\prime}_{\mu\nu]}-3A^{s}_{[\lambda}B^{s}_{\mu\nu]}.
\nonumber
\end{eqnarray}
5-d metric is given in terms of 4-d metric as
\begin{equation}
ds^{2}=e^{2\sigma}(dx^{5}+A_{t}dt)^{2}+g_{\alpha\beta}dx^{\alpha\beta} .
\end{equation}
Setting $\varphi =2\phi-\sigma$ and using the Einstein metric
$ g_{\alpha\beta}^{E}=e^{-\varphi} g_{\alpha\beta}$, we obtain
\begin{eqnarray}
S_{4}^{\prime}&=&\hat{\kappa}_{0} \int \sqrt{-g_{E}}\, d^4x\, (R_{E}-
(\nabla \sigma)^2
 - \frac{1}{2} (\nabla \varphi)^2-\frac{1}{12}e^{-2\varphi}
(\hat{H}_{\mu\nu\lambda})^2  \\
  & &-\frac{1}{4}e^{-2\sigma-\varphi} (B_{\mu\nu}^{s})^2
-\frac{1}{4}e^{2\sigma-\varphi} (F_{\mu\nu}^{s})^2 +o(\alpha^{\prime})).
\nonumber
\end{eqnarray}
This is equal to $ S_{2}$.

Using the relation~(\ref{rel:1}) we can see that the black hole solutions
for the 6-d Kaluza Klein supergravity theory are the solutions of
$S_{4}^{\prime}$
with $\hat{H}_{\mu\nu\lambda}=0$ if expressed in the new variables.
For later use we present a black hole solution. It is given by
\begin{eqnarray}
ds^{2}_{E}&=&-\frac{r-r_{+}}{\sqrt{r^{2}-r_{0}^{2}}}dt^{2}
 +\frac{\sqrt{r^{2}-r_{0}^{2}}}{r-r_{+}}dr^{2}
+(r-r_{-})\sqrt{r^{2}-r_{0}^{2}} \, d\Omega^{2}, \label{eq:n4}
\end{eqnarray}
\begin{eqnarray}
\exp\varphi = \frac{r-r_{-}}{\sqrt{r^{2}-r_{0}^{2}}} , & &
\exp\sigma = \sqrt{\frac{r+r_{0}}{r-r_{0}}} , \nonumber \\
A_{t}= -\frac{Q}{r+r_{0}} , & &
B_{t}= -\frac{P}{r-r_{0}} , \nonumber
\end{eqnarray}
 with
\begin{eqnarray}
(r_{+}-r_{0})(r_{-}-r_{0})&=& P^{2},  \\
(r_{+}+r_{0})(r_{-}+r_{0})&=& Q^{2} . \nonumber
\end{eqnarray}
Here we set $\varphi_{\infty}=\sigma_{\infty}=0$.
For a extremal black hole, we can use the expressions ~(\ref{eq:m1}) and
{}~(\ref{eq:m2}). The result is
\begin{eqnarray}
ds_{E}^{2}=-\frac{\rho}{\sqrt{(\rho+|P|)(\rho+|Q|)}}dt^{2}
+\sqrt{(1+\frac{|P|}{\rho})(1+\frac{|Q|}{\rho})}
(d\rho^{2}+\rho^{2}d\Omega^{2}),  \label{eq:n5}
\end{eqnarray}
\begin{eqnarray}
e^{\varphi}= \frac{\rho}{\sqrt{(\rho+|P|)(\rho+|Q|)}},  & &
e^{\sigma}=\sqrt{\frac{\rho+|Q|}{\rho+|P|}} , \nonumber \\
A_{t}^{\prime}=\frac{\rho}{\rho+|Q|},  & &
B_{t}^{\prime}=\frac{\rho}{\rho+|P|} . \nonumber
\end{eqnarray}
In order to obtain ~(\ref{eq:n5}) from ~(\ref{eq:n4}), we can use
$r_{+}-r_{0}=|P|$ and $r_{+}+r_{0}=|Q|$ for extremal case and define
$r-r_{+} \equiv \rho$ and we recover ~(\ref{eq:n5}). But we need a constant
shift of the vector potential which is a gauge transformation.
Concretely,
\begin{equation}
A_{t}^{\prime}-A_{t}=\frac{\rho}{\rho+|Q|}+\frac{Q}{\rho+r_{+}+r_{0}}
=\frac{\rho+Q}{\rho+|Q|}=1
\end{equation}
if $Q\geq 0$. And similarly $B_{t}^{\prime}=B_{t}+1$ if $P\geq 0$.

 The multi black hole solutions found at ~(\ref{eq:m1}) and ~(\ref{eq:m2})
 are transformed into
the solutions found by Horowitz and Tseytlin. ((4.6) in their
paper\cite{H-T}.)
Those solutions are exact solutions to all orders in
$\alpha^{\prime}$. These bosonic solutions are also shown to be exact
in the closed superstring and in the heterotic string theory as well.

For one black hole configuration in 4-dimensions,  the corresponding 5-metric
is
\begin{eqnarray}
ds^{2} &=& \frac{\rho+|Q|}{\rho+|P|}(dx^{5}+\frac{\rho}{\rho+|Q|}dt)^{2}
-\frac{\rho^{2}dt^{2}}{(\rho+|P|)(\rho+|Q|)}+d\rho^{2}+\rho^{2}d\Omega^{2} \\
      &=&  \frac{\rho+|Q|}{\rho+|P|}(dx^{5})^{2}+\frac{2\rho}{\rho+|P|}
dx^{5}dt+d\rho^{2}+\rho^{2}d\Omega^{2} . \nonumber
\end{eqnarray}
Here the internal coordinate $x^{5}$ becomes  a null coordinate in this
extremal limit and the 5-dimensional
metric describes the gravitational plane wave. If $|P|=0$, the metric
describes the usual 5-d geometry of Kaluza-Klein electric black hole.
 The wavelike behavior can be understood in the following heuristic way.
One can obtain this solution by boosting the Schwarzschild solution into
the fifth
dimension. This still satisfies the 5-dimensional field equation. However
when reinterpreted in 4-dimension, one obtains a solution with a nonzero
Maxwell field and dilaton. These solutions are regular if the velocity in the
fifth direction is subliminal. As it tends to that of light, they become
singular. In this limit, after simultaneously rescaling the size we obtain
the above solution with $|P|=0$, which describe pointlike singularities
moving with the velocity of light. It is not clear whether a similar
interpretation is possible for $|P||Q| \neq 0$.

\section{\large\bf  Rotating black hole solutions of the 5-dimensional
string theory
compactified into 4-dimensions}

 There is a close relationship between the 5-dimensional string theory
compactified into 4-dimensions, and 4-dimensional heterotic string theory
with toroidal compactification. If we define $\Phi=2\phi_{s}-\sigma$, then
$\tilde{S}_{4}$ in ~(\ref{eq:n3}) is written as
\begin{eqnarray}
\tilde{S}_{4} &=& \int d^{4}x\sqrt{-g}e^{-\Phi}(R+(\partial_{\mu}\Phi)^{2}
-(\partial_{\mu}\sigma)^{2}-\frac{1}{4}e^{2\sigma}(F^{s}_{\mu\nu})^{2}
-\frac{1}{4}e^{-2\sigma}(B^{s}_{\mu\nu})^{2}) \label{eq:SS4}
\end{eqnarray}
On the other hand, the massless fields in heterotic string theory
compactified on a six dimensional torus consists of the metric $g_{\mu\nu}$,
the antisymmetric tensor field $B_{\mu\nu}$, 28 $U(1)$ gauge fields
$A_{\mu}^{(a)}$ ($1 \leq a \leq 28$), the scalar dilaton field $\Phi$, and a
$28 \times 28$ symmetric matrix valued scalar field $M$ satisfying,
\begin{eqnarray}
MLM^{T}=L, &M^{T}=M.
\end{eqnarray}
Here $L$ is a $28 \times 28 $ symmetric matrix with 22 eigenvalues $-1$ and 6
 eigenvalues $ +1$. For definiteness we will take $L$ to be
\begin{equation}
L= \left( \begin{array}{cc}
 -I_{22} &  \\
    & I_{6} \end{array} \right),
\end{equation}
where $I_{n}$ denotes an $n \times n$ identity matrix. The action describing
the effective field theory of these massless bosonic fields is given by
\cite{Sen3},
\begin{eqnarray}
S&=& C\int d^{4}x\sqrt{-g}e^{-\Phi}(R+(\partial_{\mu}\phi)^{2}
+\frac{1}{8}Tr(\partial_{\mu}ML\partial^{\mu}L)-F_{\mu\nu}^{(a)}
(LML)_{ab}F^{\mu\nu(b)}   \label{eq:SS5} \\
 & & -\frac{1}{12}H_{\mu\nu\rho}H^{\mu\nu\rho}), \nonumber
\end{eqnarray}
 where,
\begin{eqnarray}
F_{\mu\nu}^{(a)}&=&\partial_{\mu}A_{\nu}^{(a)}-\partial_{\nu}A_{\mu}^{(a)},
\nonumber \\
H_{\mu\nu\rho}&=&3\partial_{[\mu}B_{\nu\rho]}+6A^{(a)}_{[\mu}F_{\nu\rho]}^{(b)}
L_{ab} .  \label{eq:h}
\end{eqnarray}
If we choose the special $M$
\begin{equation}
M=\left( \begin{array}{cccc}
\cosh2\sigma & & \sinh2\sigma &   \\
    & -I_{21} &        &    \\
\sinh2\sigma & & \cosh2\sigma & \\
    &  &  & I_{5}  \end{array}   \right),
\end{equation}
then $S$ can be reduced to
\begin{eqnarray}
S_{4} &=& \int d^{4}x\sqrt{-g}e^{-\Phi}(R+(\partial_{\mu}\Phi)^{2}
-(\partial_{\mu}\sigma)^{2} \\
 & & -e^{2\sigma}(\frac{F_{\mu\nu}^{(1)}-F_{\mu\nu}^{(23)}}{\sqrt{2}})^{2}
-e^{-2\sigma}(\frac{F_{\mu\nu}^{(1)}+F_{\mu\nu}^{(23)}}{\sqrt{2}})^{2}).
\nonumber
\end{eqnarray}
Thus if we set
\begin{eqnarray}
F^{s}_{\mu\nu}&=&\sqrt{2}(F_{\mu\nu}^{(1)}-F_{\mu\nu}^{(23)}) , \\
B^{s}_{\mu\nu}&=&\sqrt{2}(F_{\mu\nu}^{(1)}+F_{\mu\nu}^{(23)}),  \nonumber
\end{eqnarray}
we recover the action ~(\ref{eq:SS4}). Hence we can read off the black hole
solutions of ~(\ref{eq:SS4}) from Sen's results, which construct the
general electrically charged rotating black hole solutions of the heterotic
 string theory.

But there is a little difference in the definition of the antisymmetric
tensor fields between Sen's work and Horowitz and Tseytlin's.
 Following the Sen's definition ~(\ref{eq:h}),  we obtain
\begin{equation}
H_{\mu\nu\lambda}=3\partial_{[\mu}B_{\nu\lambda]}
-\frac{3}{2}A_{[\mu}^{s}B^{s}_{\nu\lambda]}
-\frac{3}{2}B_{[\mu}^{s}F^{s}_{\nu\lambda]},
\end{equation}
while Horowitz and Tseytlin use
\begin{equation}
\hat{H}_{\mu\nu\lambda}=3\partial_{[\mu}B^{\prime}_{\nu\lambda]}
-3A_{[\mu}^{s}B^{s}_{\nu\lambda]}.
\end{equation}
If we set $B^{\prime}_{\mu\nu}=B_{\mu\nu}-A^{s}_{[\mu}B^{s}_{\nu]}$,
$H_{\mu\nu\lambda}$ is equal to $\hat{H}_{\mu\nu\lambda}$. Thus two
definitions differ by field redefinition and this redefinition does not
change the gauge invariant field strength. One can check that equations of
motion do not change under the above field redefinition.
The rotating black hole solutions are given by
\begin{eqnarray}
ds_{E}^{2}&=& -\frac{\rho^{2}+a^{2}\cos^{2}\theta-2m\rho}{\sqrt{\Delta}}dt^{2}
+\frac{\sqrt{\Delta}}{\rho^{2}+a^{2}-2m\rho}d\rho^{2}+
\sqrt{\Delta}d\theta^{2} \nonumber  \\
 & & +\frac{\sin^{2}\theta}{\sqrt{\Delta}}[\Delta+
a^{2}\sin^{2}\theta(\rho^{2}+a^{2}\cos^{2}\theta+2m\rho \cosh\alpha
\cosh\beta)]
d\phi^{2}  \nonumber \\
 & & -\frac{2}{\sqrt{\Delta}}m\rho a \sin^{2}\theta(\cosh\alpha+\cosh\beta)
dtd\phi,
\nonumber
\end{eqnarray}
where $ds_{E}^{2}=e^{-\Phi}g_{\alpha\beta}dx^{\alpha}dx^{\beta}$ is the
Einstein metric and
\begin{eqnarray}
\Delta&=&[\rho^{2}+a^{2}\cos^{2}\theta+m\rho(\cosh\alpha \cosh\beta-1)]^{2}
-m^{2}\rho^{2}\sinh^{2}\alpha \sinh^{2}\beta.
\end{eqnarray}
The other fields are given by
\begin{eqnarray}
\exp{\Phi}&=&\frac{\rho^{2}+a^{2}\cos^{2}\theta}{\Delta} \\
\exp{\sigma}&=& \frac{\rho^{2}+a^{2}\cos^{2}\theta+
m\rho(\cosh\alpha \cosh\beta-\sinh\alpha \sinh\beta-1)}{\Delta} \nonumber
\end{eqnarray}
\begin{eqnarray}
A_{t}^{s}&=& -\Delta^{-1}m\rho [(\rho^{2}+a^{2}\cos^{2}\theta)
(\cosh\beta \sinh\alpha-\cosh\alpha \sinh\beta) \\
 & &+ m\rho(\cosh\alpha-\cosh\beta)(\sinh\alpha+\sinh\beta)] \nonumber
\end{eqnarray}
\begin{eqnarray}
A_{\phi}^{s}&=& \Delta^{-1}m\rho a \sin^{2}\theta[(\rho^{2}+
a^{2}\cos^{2}\theta)
(\sinh\alpha-\sinh\beta)  \nonumber\\
& & +m\rho(\cosh\alpha-\cosh\beta)(\sinh\alpha \cosh\beta+
\sinh\beta \cosh\alpha)], \nonumber
\end{eqnarray}
\begin{eqnarray}
B_{t}^{s}&=& -\Delta^{-1}m\rho[(\rho^{2}+a^{2}\cos^{2}\theta)
(\cosh\beta \sinh\alpha +\cosh\alpha \sinh\beta) \\
  & & +m\rho (\cosh\alpha-\cosh\beta)(\sinh\alpha-\sinh\beta)], \nonumber
\end{eqnarray}
\begin{eqnarray}
B_{\phi}^{s}&=& \Delta^{-1}m\rho a \sin^{2}\theta[(\rho^{2}+
a^{2}\cos^{2}\theta)
(\sinh\alpha+\sinh\beta)  \nonumber\\
  & & +
m\rho(\cosh\alpha-\cosh\beta)(\sinh\alpha \cosh\beta-
\sinh\beta \cosh\alpha)], \nonumber
\end{eqnarray}
\begin{eqnarray}
B_{t\phi}&=& \Delta^{-1}m\rho a \sin^{2}\theta(\cosh\alpha-\cosh\beta)
[(\rho^{2}+a^{2}\cos^{2}\theta)+m\rho(\cosh\alpha \cosh\beta -1)], \nonumber
\end{eqnarray}
and  $B^{\prime}_{\mu\nu}=B_{\mu\nu}-A^{s}_{[\mu}B^{s}_{\nu]}$.
 The various properties of the above solutions are studied at \cite{Sen3}.
Non-extremal solutions with non-zero angular momentum have two horizons
and their global structure are similar to that of the Kerr black hole
solutions. The extremal limit with non-zero angular momentum has non-zero
surface area and zero temperature. When $a=0$, the above solution
 describes spherically symmetric black holes. If we set
\begin{eqnarray}
 r \equiv \rho+m(\cosh\alpha \cosh\beta -1),  &
 r_{+} =m(1+\cosh\alpha \cosh\beta)  \\
r_{-}=m(cosh\alpha cosh\beta -1), &r_{0} = -m \sinh\alpha \sinh\beta
 \nonumber \\
Q=m(\sinh\alpha \cosh\beta-\sinh\beta \cosh\alpha), &
P=m(\sinh\alpha \cosh\beta+\sinh\beta \cosh\alpha) \nonumber
\end{eqnarray}
we obtain the previous solution ~(\ref{eq:n4}).

\section{\large\bf  Discussion}

We started with black hole solutions of 6-d Kaluza-Kline theory and found
interesting connections of those solutions with string theories. Two
kinds of string theories are mainly discussed. One is 5-d bosonic string
theory compactified into 4-dimensions, and the other is 4-d heterotic
string theory with toroidal compactification. Actually, black hole solutions
of 6-d Kaluza-Klein theory and of 5-d string theory can be read off from those
 of the heterotic string theory. Black hole solutions of 5-d
string theory can also be embedded into the closed superstring theory.
It is not surprising that such connection between  Kaluza-Klein black hole
solutions and string theories.
Many of supergravity theories can be obtained by dimensional reduction
of underlying supergravity theory of the closed superstring theory or the
heterotic string theory with consistent truncation of some fields.
Thus the embedding of massless fields of 5-d string theory can be regarded as
the embedding of the underlying supergravity theory into string theories.

Once black hole solutions of 5-d string theory is embedded into
the heterotic string theory or the closed string theory, we can generate
other solutions using T-duality transformations. However, it's not clear that
the transformed solutions are also exact solutions of the underlying
string theory since the leading order duality transformation can receive
$\alpha^{\prime}$ corrections. It remains to be seen if a similar argument
of exactness can be given to the transformed solutions as Horowitz and
Tseytlin
did.

As this work is competed, we receive a preprint by M. Cvetic\v{c} et. al.
\cite{C2} which worked out non-extremal solutions of 6-d Kaluza-Klein theory
independently. M. Cveti\v{c} informed the author that the non-extremal
solutions were
presented in another preprint\cite{C3} in December, 94. But this preprint
was not submitted to the hep-th bulletin board. The author thanks to her
for sending this preprint to him.

{\large\bf Acknowledgment}

It is my pleasure to thank A. Dabholkar, J. Gauntlett, J. Preskill, and
J. Schwarz for the useful discussions. I would like also to thank P. Yi for
the relevant suggestions on the paper and J. Preskill for reading the
manuscript.

\newpage

\noindent
{\bf Figure Caption}

\noindent
Fig.1: Penrose diagrams for the black hole solutions.
Fig.1a is for the non-extremal case, Fig.1b is for extremal black holes
with both charges non zero and Fig.1c corresponds to extremal black holes with
only one charge.

\end{document}